\documentclass{emulateapj}

\shorttitle{Pulsation coherence in XTE J1814-338}
\shortauthors{Watts, Patruno \& van der Klis}

\begin{document}

\title{Coherence of burst oscillations and accretion-powered pulsations in the accreting millisecond pulsar XTE J1814-338}

\author{Anna L. Watts\altaffilmark{1}, Alessandro Patruno \& Michiel van der Klis} 
\affil{Astronomical Institute ``Anton Pannekoek'', University of
  Amsterdam, Kruislaan 403, 1098 SJ Amsterdam, the Netherlands}
\altaffiltext{1}{A.L.Watts@uva.nl}

\begin{abstract}
X-ray timing of the accretion-powered pulsations during the 2003
outburst of the accreting millisecond pulsar XTE J1814-338 has revealed
variation in the pulse time of arrival residuals. These can be interpreted in several ways, including spin-down and wandering of the fuel impact point around the magnetic pole. In this Letter we show that the 
burst oscillations of this source are coherent with the
persistent pulsations, to the level where they track all of the
observed fluctuations.  Only one burst, which occurs at
the lowest accretion rates, shows a significant phase offset.  We
discuss what might lead to such rigid phase-locking between the modulations in the accretion and thermonuclear burst emission, and consider the implications for spin variation and
the burst oscillation mechanism. Wandering of the fuel impact hot spot
around a fixed magnetic pole seems the most likely cause for the accretion-powered
pulse phase variations. This means that the burst asymmetry is coupled
to the hot spot, not the magnetic pole.  If premature ignition at this point (due to higher local
temperatures) triggers a burning front that stalls before
spreading over the entire surface, the resulting localized nuclear hot
spot may explain the unusual burst and burst oscillation properties of this source.
\end{abstract}


\section{Introduction}

The accreting millisecond X-ray pulsars (AMXPs) are a small class of  neutron stars in
Low Mass X-ray Binaries that show pulsations in outburst,
thought to be caused by magnetic channeling of accreting plasma.  Detailed timing studies of these stars reveal
 diverse behavior that can be interpreted in terms of
 spin variation or shifts in emission pattern. Both processes may play a role, and the
degree to which we can be confident in inferred values of 
spin-up or spin-down remains a hot topic \citep{gal02, bur06, bur07, 
  pap07, pap08, har08, rig08}.  

Some additional way of verifying the
timing analysis obtained from the accretion-powered pulsations would be useful, and in
this respect two of the AMXPs are particularly valuable.  SAX
J1808.4-3658 (J1808) and XTE J1814-338 (J1814)
also show thermonuclear-powered
pulsations, or burst oscillations.  These are high frequency 
variations seen during Type I X-ray bursts, powered by unstable
burning of accreted fuel.  In these systems the burst oscillation
frequency is at (J1814), or very close to (J1808), the spin frequency
(\citealt{cha03, str03}, hereafter S03). The frequency is stable in the decaying tails of the bursts, with
 no sign of the large frequency drifts seen in other burst oscillation
 sources \citep{mun02}.  In J1814 the frequency is also stable during
 the rising phase of the bursts, making it the most straightforward
 candidate for burst oscillation timing (\citealt{wat05}, hereafter
 W05).   

J1814's accretion-powered pulsations show significant pulse time of
arrival (TOA) residuals even after correction for orbital Doppler
shifts (\citealt{pap07}, hereafter P07).  The cause is still a matter
of debate:  P07 interpreted the observations in terms of a steady
spin-down coupled with some jitter due to wandering of the fuel impact
hotspot around the magnetic pole.  However there are other
possibilities, such as changes in beaming due to the accretion shock,
that may also lead to the observed variation.  In this respect
analysis of the burst oscillations may be simpler:  although the
process responsible is not yet understood, the thermal spectrum
suggests a purely surface mechanism, with the accretion shock
contributing little to the observed asymmetry.   

Some initial investigation of this issue was carried out by S03, who
found that burst oscillations in the first 12 bursts were phase-locked
to within 2.5$^\circ$ of the persistent pulsations. Their analysis, however, covered only the
first 10 days of the $\approx$ 50 day outburst; before any of the variation reported by
P07 is apparent.  The level of coherence between the two types of
pulsations is also important in our efforts to understand the burst
oscillation mechanism, which remains mysterious (see the
reviews by \citealt{str06}, \citealt{gal06}).  The AMXPs are  
the only sources in which we can quantify the role of the magnetic
field. We want to know, for example,  whether the burst oscillations couple to
  the magnetic field or to the fuel stream impact point.

\section{Timing analysis}
\label{data}

J1814 was discovered in 2003 in the  {\it Rossi
X-ray Timing Explorer (RXTE)} Galactic bulge monitoring campaign \citep{mar03a}, and
remained in outburst for nearly 2 months.  The pulsar has a spin
frequency of 314.4 Hz and resides in a binary with an orbit of 4.3 hours
\citep{mar03}.  During the outburst over 425 ks of high time
resolution data were taken with {\it RXTE}'s Proportional Counter Array
(PCA, \citealt{jah06}).  A total of 28 X-ray bursts were detected, all with burst
oscillations at the spin frequency. Both accretion-powered pulsations
and burst oscillations have a strong overtone at twice the spin
frequency (S03).   

For our timing analysis we use all available pointed observations from
the {\it RXTE} PCA with Event mode data (time resolution 122
$\mu$s, 64 binned energy channels) or Good Xenon mode data (time resolution 1 $\mu$s, 256 unbinned energy channels).  The latter
were rebinned in time to 122 $\mu$s time resolution.  Data were
barycentered using the JPL DE405 
ephemeris and a spacecraft ephemeris including fine clock
corrections which together provide an absolute timing accuracy of 3-4
$\mu$s \citep{rot04}, using the source position of
\citet{kra05}.

In analysing the accretion-powered pulsations we discard the X-ray
bursts, removing all data from 50 s before to 200 s after the burst
rise, and select only photons in the 2.5-17 keV range to maximize the
signal to noise ratio.  Pulse profiles
are built using the fixed frequency solution of P07
to fold segments of approximately 500 s of data
after subtracting the background contamination (using the FTOOL
{\it{pcabackest}}).  The TOAs were then obtained by cross-correlating the folded profiles
with a pure sinusoid whose frequency represents the spin frequency of
the neutron star. The same procedure is repeated for the first
overtone.  The fiducial point used in measuring the TOAs was
the peak of the sine wave being cross-correlated.  The determination of
pulse TOAs and their statistical uncertainties follows the
standard radio pulsar technique \citep{tay93}.  Fitting a
Keplerian orbit plus a constant spin frequency $\nu$, or spin
derivative $\dot{\nu}$, we obtain solutions consistent with those of
P07.  

Like P07, we find a
strong anti-correlation between flux and the accretion-powered
pulse TOA residuals. It is interesting, however, that this
anti-correlation becomes weaker when one considers the residuals from
an ephemeris that includes a constant $\dot{\nu}$ term.  A rank
correlation test between flux and residuals 
for a constant frequency model gives a Spearman
coefficient of $\rho=-0.71$ (fundamental) and $\rho=-0.90$ (first
overtone), with a probability of $< 0.01$\% that
the two variables are not anti-correlated.  Including spin-down, the
magnitudes of the
Spearman coefficients fall to $\rho=-0.56$ (probability still $<
0.01$\% that the two variables are not anti-correlated, but larger
than the probabilities for the zero spin derivative case) and
$\rho=-0.12$ (a probability of 27\% that the two variables are
not anti-correlated) respectively. As we will argue in Section \ref{disc}, an
accretion-rate dependent hot spot location may be able to explain the entire
residual record, with no need for spin variation.  

\begin{figure}
  \centering
    \includegraphics[width=8.5cm, clip]{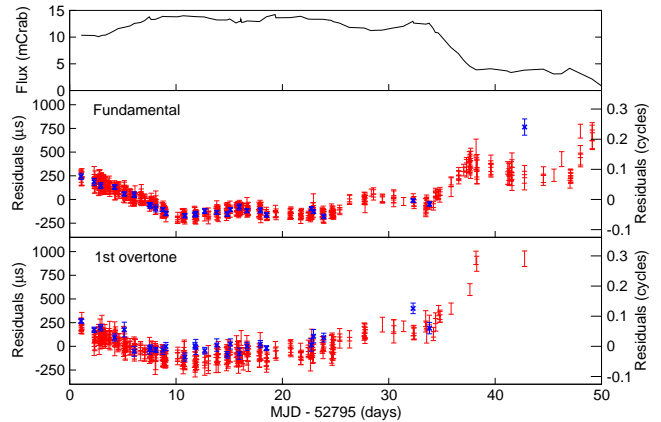}
     \caption{Phase residuals for the accretion-powered pulsations (bars; red) and the
     burst oscillations (crosses; blue), compared to a model that has a
     constant frequency and orbital Doppler shifts.  The final burst,
     which has no detectable 1st overtone component, requires special
     treatment; see text. The units on the right axis are {\it rotational}
     cycles.  The top panel shows the outburst light curve. }
     \label{mult}
 \end{figure}

We then apply the same timing procedure to the X-ray bursts.  As in W05 and
\citet{wat06} we use data where the count rate is at least twice 
the pre-burst level.  For the first 27 bursts there is no evidence of frequency
variability during the bursts (W05), so we use the entire burst to
generate a folded profile.  The final burst requires 
more care, as there is a statistically significant frequency drop in
the late stages of the burst rise (Fig.20 of W05). For this burst we use
only the first 2 s of the burst rise, before the frequency starts to
shift. For each folded burst profile we then compute residuals using
the same ephemeris that we used for the accretion-powered pulsations.
Again we cross-correlate the folded profile using both a fundamental
and an overtone.  Figure \ref{mult} shows the TOA residuals for the accretion-powered
pulsations and the burst oscillations.

To check whether the TOAs of the burst oscillations are consistent
with having the same temporal dependence as the accretion-powered
pulsations we fit a constant frequency model to the two TOA sets
separately (excluding the final burst, see below).  The fitted frequency is the
same within the statistical uncertainties both for the fundamental and
the first overtone. We also tried fitting the two TOA sets
with a spin frequency plus a frequency derivative.  Again the two
solutions are consistent within the statistical uncertainties.
However P07
already noted that both ephemerides are a poor fit to the
accretion-powered pulse TOA residuals, and the same is true for the
burst oscillation residuals.  They
are formally inconsistent with a constant frequency model: for the fundamental
this assumption gives a $\chi^{2}$ of 459 for 23 degrees of freedom
(dof). They are also inconsistent with a constant $\dot{\nu}$, as
$\chi^{2}$ in this case is still large, 80 for 22 dof. Similar
results are obtained for the first overtone.  

It is clear from the Figure that, for all but the final burst, the burst
oscillation TOA residuals track the accretion pulsation      
residuals.  To test the phase-lock we computed the phase difference $\Delta
\phi_m = \nu[{\rm TOA}_{\rm bur} - {\rm TOA}_{\rm acc}]$, using the accretion 
pulse TOA residuals from the observation containing each burst.  We
then fitted a constant. For
the fundamental we found a $\chi^2$ of 15.3 (26
dof), with best fit $\Delta \varphi_m =  (0.004 \pm 0.002)$ rotational cycles.
For the first overtone we found a $\chi^2$ of 11.3 (26 dof), with
best fit $\Delta \varphi_m = -(0.001 \pm 0.003)$ rotational cycles. The fact that a constant
$\Delta \varphi_m$  is such a
good fit confirms that the two sets of
pulsations are phase-locked. The small ($2\sigma$) non-zero offset in the fundamental
 bears comment.  Unlike the burst oscillations, the
accretion-powered pulsations have soft lags across the 2.5-20 keV
band, with higher energy photons arriving earlier in phase
\citep{wat06}.  Taking this into account, we find that the burst
oscillation TOAs are completely coincident with the softer (2.5 - 5
keV) component of the 
accretion pulsations, thought to originate from
stellar surface (Section \ref{disc}).   The only exception to the
phase-locking is the final burst, with $\Delta \varphi_m = (0.2 \pm
0.04)$ rotational cycles.

When calculating burst oscillation residuals some care is required,
since if accretion continues during the burst, there might still be a contribution from the accretion-powered
pulsations.  The resulting bias in the burst
oscillation phase can be
calculated easily by considering the profile that results from the
addition of two offset sinusoidal profiles. Standard trigonometric
identities yield a relation between $\Delta \varphi_m$ (the measured
offset between burst oscillation phase and accretion-powered
pulse phase) and $\Delta \varphi_b$ (the bias, i.e. the offset between
measured and true burst
oscillation phase caused by residual accretion):  

\begin{equation}
\tan \Delta \varphi_m = \frac{\sin(\Delta\varphi_m -
  \Delta\varphi_b)}{\left[N_\mathrm{acc} r_\mathrm{acc}/N_\mathrm{bur}
    r_\mathrm{bur}\right] + \cos(\Delta\varphi_m -
    \Delta\varphi_b)}.
\label{offset}
\end{equation}
where $N_\mathrm{acc}$ and $N_\mathrm{bur}$ are the number of
accretion and burst photons in the folded profile, 
$r_\mathrm{acc}$ and $r_\mathrm{bur}$ being the associated fractional
amplitudes. The quantity $\Delta\varphi_m$ was measured earlier: $\Delta\varphi_m \lesssim 0.01$ cycles
at the 99\% confidence level.  Using the values of $N_\mathrm{acc}$,
$N_\mathrm{bur}$, $r_\mathrm{acc}$ and $r_\mathrm{bur}$ from Table 1
of W05, Equation (\ref{offset}) gives a 99\% confidence upper limit on the
bias introduced by any residual accretion pulsation of $10^{-3}$
cycles.  This is sufficiently small that it does not  
affect our analysis. Similar conclusions can be reached for the
overtone.  Note that in computing these limits we assume
that the
accretion flow parameters ($N_\mathrm{acc}, r_\mathrm{acc},
\Delta\varphi_m$) are unchanged during a burst.  If accretion is
inhibited during a burst, due to radiation pressure, the bias will be lower.

\section{Discussion}
\label{disc}

J1814 has unusual burst oscillation properties
compared to other sources: they occur in hydrogen-rich bursts, have
negligible frequency and amplitude variation, and have a soft spectrum
(S03, W05, \citealt{wat06}).  Our analysis has shown that the burst
oscillations are also phase-locked to the accretion-powered pulsations (to within
$3^\circ$ at 99\% confidence) despite the substantial phase-wander
exhibited by the latter over the course of the outburst.

In fact the burst oscillations are not only phase-locked but also
coincident, having the same phase as the soft (2.5-5 keV), lagging, part of the  
accretion-powered pulsations (although they are also at the 2$\sigma$ level
coincident with the entire 2.5-20 keV accretion-powered pulse). Detailed modeling of the accretion
pulsations has yet to be done for
J1814, but modeling for other AMXPs with similar pulse properties
suggests that the soft pulsed component comes from a hot spot on the stellar surface, with the hard component
originating in the accretion funnel \citep{pou03, gie05, fal07}.  

We
 first consider what the exceptional degree of phase-locking
 implies about the cause
of the variation in the TOA residuals.  There
are several parts of the system whose variation might affect both types of
pulsation: surface rotation, the accretion funnel/disk, the magnetic field, and the fuel deposition
footprint.  

{\it Case 1: Genuine changes in the spin rate of the stellar
  surface.} Our result would be consistent with a model where all of
the variation is due to spin changes, both sets
of pulsations being locked to the surface.  However this
requires alternating spin-up and spin-down with $|\dot{\nu}| \sim
10^{-12}$ Hz/s. Even if the crust were decoupled from the
fluid core, this is high compared to what is
achievable from the 
expected accretion or
gravitational wave torques \citep{and05, bil98b}.  Fitting a constant spin-down term
$\dot{\nu} \approx 6\times 10^{-14}$ Hz/s, as argued by P07, does
improve the quality of the fits somewhat.  However spin derivatives $|\dot{\nu}| \sim
10^{-12}$ Hz/s would still be required on shorter timescales to explain the
remaining excursions.  Our results would also be consistent with
precession, but modeling by \citet{chu08} suggests that
precession is unlikely in this source.

{\it Case 2: Changes in beaming/scattering by the
  accretion funnel or disk.} 
 The accretion shock in the funnel is thought to contribute to the pulsed emission of the
  accretion-powered pulsations, leaving a signature of hard emission in the
  spectrum.  If the funnel were to have a similar effect on the much
  stronger burst emission it would have to do this without leaving any
  trace in the spectrum \citep{str03, kra05}.  This does not seem
  feasible.  Our result also casts doubt on the accretion disk being
  the source of the soft lagging component of the accretion-powered
  pulsations (one of the possibilities considered by \citealt{fal07}), since it is hard to understand why the
burst oscillations (a surface process) would track a component
generated in the surrounding environs.  

{\it Case 3:  Wander of the magnetic pole, or changes in field
  geometry.}  Motion of the magnetic pole would affect
location of the accretion hot spot.  If the magnetic field also
determines the location of the nuclear burning hot spot through
modulation of ignition or emission, this could also explain our
result. However, the observed variability would require
localized burial or amplification of the poloidal field component on timescales
of order a day. The accretion rate in the peak of the outburst is at
most a few percent of the Eddington rate \citep{gal04}. Current
modeling suggests that this is insufficient to cause burial of the
polar 
field on the required timescales \citep{bro98, cum01, pay04}. 
There is also no obvious mechanism for field amplification:
material arriving via a funnel flow will have almost no angular
momentum differential compared to the stellar surface.  The heating
associated with accretion could bring a buried field to the
surface \citep{cum01}, but as previously stated burial is unlikely at
such low accretion rates.

{\it Case 4: Wander of the fuel deposition point around the magnetic
  pole.} Simulations of funneled accretion have shown that
the fuel deposition point can exhibit phase excursions around
a fixed magnetic pole as accretion rate fluctuates, particularly for small
misalignment angles between the magnetic pole and the spin axis
\citep{rom03, rom04, lam08}\footnote{The pulse profile modeling for this
  source which has been attempted has suggested large
  misalignment angles \citep{bha05, 
    lea08}.  If this were the case, the observed phase variability would require
  the fuel impact point to migrate back and forth by several km over the course
  of the outburst -  an uncomfortably large amount. It seems more likely
  that some of the assumptions in these models, particularly the use
  of single temperature circular hotspots (known to be problematic,
  see \citealt{wat06}), need to be revisited.}.  Such a model might
neatly account for the
correlation between residuals and flux without requiring any non-zero $\dot{\nu}$, since the stable position for
the fuel impact point will vary in azimuth depending on accretion
rate. The fact that the anti-correlation between
flux and residuals is stronger when we set $\dot{\nu}=0$ supports this idea.

If the most plausible explanation for the variability in the two sets of
pulsations is the last, what physical
mechanisms might lead to phase-locking between the fuel impact point -
which moves 
with accretion rate relative to field 
geometry on timescales of a few days - and the nuclear burning hot
spots?  

One possibility is some degree of magnetic
confinement, leading to accumulation of fuel at the accretion hot 
spot. Material deposited near the polar cap will be prevented from spreading until the
over-pressure is sufficient to distort the field lines \citep{bro98},
even if the impact point is not precisely on the polar cap.  To ensure that the magnetic propeller effect does not
disrupt accretion, the magnetic field of J1814 must be $\lesssim 10^9$ G
\citep{psa99, rap04}.  This is consistent with the upper
limit on spin-down inferred from  pulse timing (P07).   At this
upper limit helium could be confined until a column depth $\sim 10^6$ g
cm$^{-2}$, but this is still well before the material reaches the
bursting layer at column depth $\sim 10^8$ g cm$^{-2}$
\citep{bil98}.  Hydrogen will spread even more easily.  Based on these
estimates we conclude that fuel confinement is not effective, although
it has been advanced as a possible explanation for the short burst
recurrence times \citep{gal04}.  

The other factor distinguishing the fuel impact point is its
temperature, which is higher than the rest of the star.  The magnitude
of the temperature differential has not been 
determined observationally, but it could certainly affect burst
emission. One possibility is that the higher temperatures modify the
local composition (via steady burning between bursts, for example).
The higher starting temperature and/or composition could in principle
modify the flux from the burst once it starts;  perhaps driving more
energetic reactions or enhancing convection.  Whether this effect
would be large enough to explain the high fractional amplitudes is,
however, not clear.   

An alternative is the effect that a higher local temperature would
have on ignition conditions.  Previous studies that concluded that
ignition would occur predominantly near the equator \citep{spi02,
  coo07} did not consider the effect of non-uniformities in
temperature.   A small increase in temperature can have
a large effect on the column depth required for ignition, with
the hotter area requiring a lower column depth \citep{bil98}.
Material at the fuel impact point could therefore reach 
ignition conditions well ahead of the rest of the star.  In this scenario
the burning front might stall before spreading across the whole star,
depending on the rate of heat transfer across the burning front. This
would result in a brightness asymmetry centered on the fuel impact point.
Premature ignition at the fuel deposition point, followed by stalling, could explain
several observational features:  the rather small black body radius of
the bursts  
\citep{gal06}, the shorter than expected burst recurrence times
\citep{gal04}, and the burst shapes, which suggest off-equatorial
ignition \citep{mau08}.  Note that off-equatorial ignition alone is not
sufficient to explain the presence of an asymmetry in the burst tail:
something else, such as stalling, is required.

Whatever the mechanism, it must fail once the accretion rate
drops (or perhaps when the burst is more energetic),
since the oscillations in the final burst are offset.  The fuel impact
footprint is by this time thought to be smaller, since the fractional
amplitude of the accretion-powered pulsations is
rising (W05).  In addition the accreted fuel has more time
to spread and equilibrate between bursts.  Both factors will act to
reduce the temperature differential between the fuel stream impact point and the rest of
the star, which would make ignition at the fuel impact point less likely.   The oscillation
properties of this final burst (substantially lower fractional 
amplitude, frequency drift) are very different to the rest of
the sample, and it is quite plausible that a different burst
oscillation mechanism operates in this final burst.  

\acknowledgments
We thank Mariano Mendez and the participants of the workshop ``A
Decade of Accreting 
Millisecond X-ray Pulsars'' for the lively discussions that prompted
and informed 
this work.


\begin{thebibliography}{99}

\bibitem[\protect\citeauthoryear{Andersson et al.}{2005}]{and05}
Andersson N. et al., 2005, MNRAS, 361, 1153

\bibitem[\protect\citeauthoryear{Bhattacharyya et al.}{2005}]{bha05}
Bhattacharyya S. et al., 2005, ApJ, 619, 483

\bibitem[\protect\citeauthoryear{Bildsten}{1998a}]{bil98b}
Bildsten L., 1998, ApJ, 501, L89


\bibitem[\protect\citeauthoryear{Bildsten}{1998b}]{bil98}
Bildsten L., 1998, in The Many Faces of Neutron Stars, eds Buccheri et
al., Kluwer Academic Publishers, p.419

\bibitem[\protect\citeauthoryear{Brown \& Bildsten}{1998}]{bro98}
Brown E.F., Bildsten L., 1998, ApJ, 496, 915

\bibitem[\protect\citeauthoryear{Burderi et al.}{2006}]{bur06}
Burderi L. et al., 2006, ApJ, 653, L133


\bibitem[\protect\citeauthoryear{Burderi et al.}{2007}]{bur07}
Burderi L. et al., 2007, ApJ, 657, 961


\bibitem[\protect\citeauthoryear{Chakrabarty et al.}{2003}]{cha03}
Chakrabarty D. et al., 2003, Nature, 424, 42

\bibitem[\protect\citeauthoryear{Chung et al.}{2008}]{chu08}
Chung C. et al., MNRAS in press, eprint arXiv:0808.3820

\bibitem[\protect\citeauthoryear{Cooper \& Narayan}{2007}]{coo07}	
Cooper R.L., Narayan R., 2007, ApJ, 657, L29

	
\bibitem[\protect\citeauthoryear{Cumming et al.}{2001}]{cum01}
Cumming A. et al., 2001, ApJ, 557, 958


\bibitem[\protect\citeauthoryear{Falanga \& Titarchuk}{2007}]{fal07}
Falanga M., Titarchuk L., 2007, ApJ, 661, 1084

\bibitem[\protect\citeauthoryear{Galloway et al.}{2002}]{gal02}
Galloway D.K. et al., 2002, ApJ, 576, L137

\bibitem[\protect\citeauthoryear{Galloway et al.}{2004}]{gal04}
Galloway D.K. et al., 2004, BAAS, 36, 954
	
\bibitem[\protect\citeauthoryear{Galloway et al.}{2006}]{gal06}
Galloway D.K. et al., ApJS in press, astro-ph/0608259

\bibitem[\protect\citeauthoryear{Gierli\'nski \& Poutanen}{2005}]{gie05}
Gierli\'nski M., Poutanen J., 2005, MNRAS, 359, 1261


\bibitem[\protect\citeauthoryear{Hartman et al.}{2008}]{har08}
Hartman J.M. et al., 2008, ApJ, 675, 1468

\bibitem[\protect\citeauthoryear{Jahoda et al.}{2006}]{jah06}
Jahoda K. et al., 2006, ApJSS, 163, 401


%
\bibitem[\protect\citeauthoryear{Krauss et al.}{2005}]{kra05}
Krauss M.I. et al., 2005, ApJ, 627, 910

\bibitem[\protect\citeauthoryear{Lamb et al.}{2008}]{lam08}
Lamb F.K. et al., eprint arXiv:0808.4159

\bibitem[\protect\citeauthoryear{Leahy et al.}{2008}]{lea08}
Leahy D.A. et al., eprint arXiv:0806.0824

\bibitem[\protect\citeauthoryear{Markwardt \& Swank}{2003}]{mar03a}
Markwardt C.B., Swank J.H., 2003, IAU Circ. 8144, 1

%
\bibitem[\protect\citeauthoryear{Markwardt et al.}{2003}]{mar03}
Markwardt C.B. et al., 2003, Atel 164.

	
\bibitem[\protect\citeauthoryear{Maurer \& Watts}{2008}]{mau08}
Maurer I., Watts A.L., 2008, MNRAS, 383, 387

\bibitem[\protect\citeauthoryear{Muno et al.}{2002}]{mun02}
Muno M.P. et al., 2002, ApJ, 580, 1048


\bibitem[\protect\citeauthoryear{Payne \& Melatos}{2004}]{pay04}
Payne D.J.B., Melatos A., 2004, MNRAS, 351, 569


\bibitem[\protect\citeauthoryear{Papitto et al.}{2007}]{pap07}
Papitto A. et al., 2007, MNRAS, 375, 971


\bibitem[\protect\citeauthoryear{Papitto et al.}{2008}]{pap08}
Papitto A. et al., 2008, MNRAS, 383, 411

\bibitem[\protect\citeauthoryear{Poutanen \& Gierli\'nski}{2003}]{pou03}
Poutanen J., Gierli\'nski M., 2003, MNRAS, 343, 1301


\bibitem[\protect\citeauthoryear{Psaltis \& Chakrabarty}{1999}]{psa99}
Psaltis D., Chakrabarty D., 1999, ApJ, 521, 332


\bibitem[\protect\citeauthoryear{Rappaport et al.}{2004}]{rap04}
Rappaport S.A. et al., 2004, ApJ, 606, 436


\bibitem[\protect\citeauthoryear{Riggio et al.}{2008}]{rig08}
Riggio A. et al., 2008, ApJ, 678, 1273


\bibitem[\protect\citeauthoryear{Romanova et al.}{2003}]{rom03}
Romanova M.M. et al., 2003, ApJ, 595, 1009


\bibitem[\protect\citeauthoryear{Romanova et al.}{2004}]{rom04}
Romanova M.M. et al., 2004, ApJ, 610, 920 


\bibitem[\protect\citeauthoryear{Rots et al.}{2004}]{rot04}
Rots A.H. et al., 2004, ApJ, 605, L129


\bibitem[\protect\citeauthoryear{Spitkovsky et al.}{2002}]{spi02}
Spitkovsky A. et al., 2002, ApJ, 566, 1018


\bibitem[\protect\citeauthoryear{Strohmayer et al.}{2003}]{str03}
Strohmayer T.E. et al., 2003, ApJ, 596, L67


\bibitem[\protect\citeauthoryear{Strohmayer \& Bildsten}{2006}]{str06}	
Strohmayer T., Bildsten L., 2006,  in Compact stellar X-ray sources,
eds Lewin W., van der Klis M., Cambridge Astrophysics Series 39,
Cambridge, UK, p.113

\bibitem[\protect\citeauthoryear{Taylor}{1993}]{tay93}
Taylor J.H., 1993, in Pulsars as Physics Laboratories, ed. R.D.Blandford (New York: Oxford Univ. Press), 117


\bibitem[\protect\citeauthoryear{Watts et al.}{2005}]{wat05}
Watts A.L. et al., 2005, ApJ, 634, 547


\bibitem[\protect\citeauthoryear{Watts \& Strohmayer}{2006}]{wat06}
Watts A.L., Strohmayer T.E., 2006, MNRAS, 373, 769


\end{thebibliography}
\end{document}